\documentclass{icrc2009}

\usepackage{graphicx}   
\usepackage{caption}    
\usepackage[font=footnotesize]{subfig} 
\usepackage{fixltx2e}
\usepackage{url}

\newcommand{\shorttitle}[1]%
{\markboth{Proceedings of the 31\MakeLowercase{$^{st}$} ICRC, {\L}\'{o}d\'{z} 2009}{#1} }
\newcommand{\etal}{\MakeLowercase{\textit{et al. }}} 

\begin{document}
\title{Monitoring solar flares with \emph{Fermi}-LAT}

\author{\IEEEauthorblockN{Giulia Iafrate\IEEEauthorrefmark{1}\IEEEauthorrefmark{2},
              Francesco Longo\IEEEauthorrefmark{2},
              Nicola Giglietto\IEEEauthorrefmark{3} and
              Monica Brigida\IEEEauthorrefmark{3}\\on behalf of the \emph{Fermi}-LAT Collaboration}
             \\
\IEEEauthorblockA{\IEEEauthorrefmark{1}INAF-Astronomical Observatory of Trieste, 
Italy} \IEEEauthorblockA{\IEEEauthorrefmark{2}INFN Trieste, Italy} 
\IEEEauthorblockA{\IEEEauthorrefmark{3} Dipartimento Interateneo di Fisica di 
Bari and INFN Bari, Italy}}

\shorttitle{G. Iafrate \etal Solar flares with \emph{Fermi}-LAT} \maketitle

\begin{abstract}
\emph{Fermi}-LAT is performing an all-sky gamma-ray survey from 20 MeV to $>$ 300 
GeV with unprecedented sensitivity and angular resolution. \emph{Fermi} is the 
only mission able to detect high energy ($>$~20 MeV) emission from the Sun during 
the new solar cycle 24. \emph{Fermi} was launched on June 2008, since then high 
energy emission from the Sun was continuously monitored searching for flare 
events. Upper limits were derived for all the solar flares detected by other 
missions and experiments (RHESSI, \emph{Fermi}-GBM, GOES). We present the 
analysis techniques used for this study and the preliminary results obtained so 
far.
  \end{abstract}


 \begin{figure*}[th]
  \centering
  \includegraphics[width=5in]{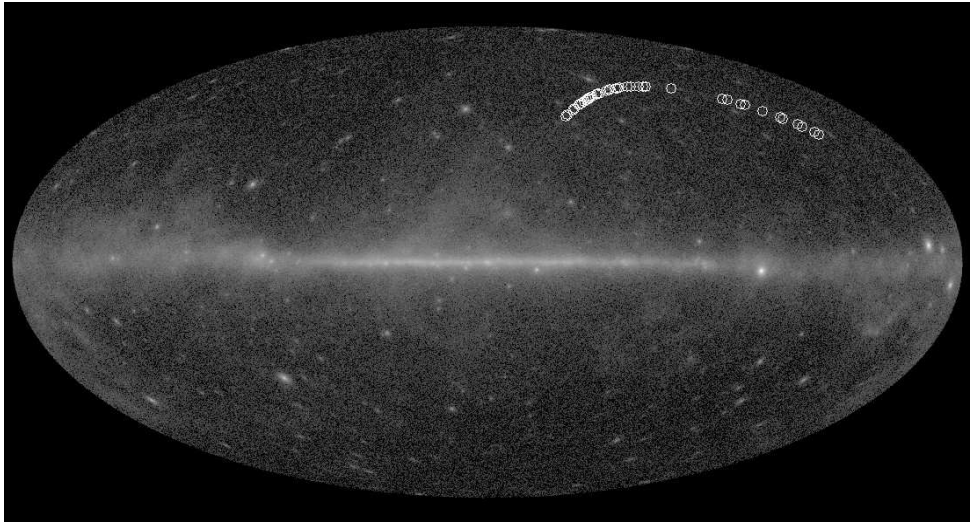}
  \caption{Location of solar flares detected by RHESSI superimposed on a count map of
  the first three months (August - October 2008) of LAT data ($E>200\textnormal{ MeV}$). There is no evidence of correlation
  between flare positions and excesses of the LAT events.}
  \label{cmap}
 \end{figure*}
 
\section{Introduction}
\emph{Fermi} was successfully launched from Cape Canaveral on 2008 June 11. It is 
currently in an almost circular orbit around the Earth at an altitude of 565 km 
having an inclination of 25.6$^{\circ}$ and an orbital period of 96 minutes. 
After an initial period of engineering data taking and on-orbit 
calibration\cite{Fermi}, the observatory was put into a sky-survey mode in August 
2008. The observatory has two instruments onboard, the Large Area Telescope 
(LAT)\cite{LAT}, a pair-conversion gamma-ray detector and tracker (energy range 
20 MeV - $>300$ GeV) and a Gamma Ray Burst Monitor (GBM), dedicated to the 
detection of gamma-ray bursts (energy range 8 keV - 40 MeV). The instruments on 
\emph{Fermi} provide coverage over the energy range measurements from few keV to 
several hundreds of GeV.\\ Here we report results of the monitor of solar flares 
seen by other missions searching for flares seen by \emph{Fermi}-LAT. Solar 
flares are the most energetic phenomena that occur within our Solar System. A 
flare is characterized by the impulsive release of a huge amount of energy, 
previously stored in the magnetic fields of active regions. During a flare plasma 
of the solar corona and chromosphere is accelerated and electromagnetic radiation 
covering the entire spectrum is emitted. The production of $\gamma$-rays involves 
flare-accelerated charged-particle (electrons, protons and heavier nuclei) 
interactions with the ambient solar atmosphere. Electrons accelerated by the 
flare, or from the decay of $\pi^{\pm}$ secondaries produced by nuclear 
interactions, yield X and $\gamma$-ray  bremsstrahlung radiation with a spectrum 
that extends to the energies of the primary particles. Proton and heavy ion 
interactions also produce $\gamma$-rays through $\pi^{0}$ decay, resulting in 
a spectrum that has a maximum at 68 MeV\cite{share}.\\
Intensity and frequency of solar flares depend on the Sun activity, according to 
the 11 year solar cycle. Most intense flares occur during the maximum, but 
intense flares can occur also in the rising and decreasing phases of the cycle. 
The new solar activity cycle 24 has started at the beginning of year 2008, the 
maximum is predicted in year 2012. \emph{Fermi} has been launched during the 
minimum of the solar cycle, so frequency and intensity of solar flares will 
increase throughout most of the mission. If the goal of a 10-year mission life is 
achieved, \emph{Fermi} will operate for nearly the entire duration of solar cycle 
24. During this time, \emph{Fermi} will be the only high-energy observatory to 
complement several solar missions at lower energies: RHESSI, GOES, SoHO.\\

\section{Previous observations} 

The 2005 January 20 solar flare produced one of the most intense, fastest rising, 
and hardest solar energetic particle events ever observed in space or on the 
ground. $\gamma$-ray measurements of the flare\cite{share06}\cite{grechnev} 
revealed what appear to be two separate components of particle acceleration at 
the Sun: i) an impulsive release lasting $\sim10$ min with a power-law index of 
$\sim3$ observed in a compact region on the Sun and, ii) an associated release of 
much higher energy particles having a spectral index $\leq2.3$ interacting at the 
Sun for about two hours. Pion-decay $\gamma$-rays appear to dominate the latter 
component. Such long-duration high-energy events have been observed before, most 
notably on 1991 June 11 when the EGRET instrument on CGRO observed $>50$ MeV 
emission for over 8 hours\cite{kanbach}. It is possible that these high-energy 
components are directly related to the particle events observed in space and on 
Earth.\\ \emph{Fermi} will improve our understanding of the mechanisms of the 
$\gamma$-ray emission by solar flares thanks to its large effective area, 
sensitivity and high spatial and temporal resolution.

\section{Monitor of solar cycle 24}

The solar cycle 24 has started at the beginning of 2008, but actually we are in 
an extended period of minimal solar activity. We are seeing an interesting 
diminished level of activity. There are some discussion ongoing if sunspots and 
flares ever return and how unusual is this behavior\cite{nugget}. A closer look 
at the daily values of three indices: F10.7 (10 cm radio flux from the Sun), the 
total solar irradiance TSI, and the classical sunspot number gives a clear 
appearance of a little up-turn. Probably F10.7 is giving us an early warning 
about the sudden increase of cycle 24 spots. These should appear within the next 
few weeks or months.\\ In the modern era there is no precedent for such a 
protracted activity minimum, but there are historical records from a century ago 
of a similar pattern (transition between cycles 13 and 14, 107 year ago). We do 
expect activity to pick up fairly suddenly soon. In the meanwhile is a good 
opportunity to use the excellent data available from many satellites to improve 
LAT analysis of solar flare  and practise in flare monitoring and analysis, to be 
ready when the first intense flare of cycle 24 will arrive.

\section{Data selection}

Since August 2008 we monitor continuously the flares detected by RHESSI and GOES, 
analysing LAT data for flare events potentially detectable by the LAT and 
computing upper limits on the solar high energy emission. We searched for solar 
flares in the LAT data from August 2008 to the end of May 2009. We analysed LAT 
data in the time intervals of flares detected by GBM, RHESSI and GOES. We have 
applied a zenith cut of $105^{\circ}$ to eliminate photons from the Earth's 
albedo. For this analysis we adopted the \lq\lq Diffuse\rq\rq\ class\cite{LAT} 
selection, corresponding to the events with the highest photon classification 
probability, using the IRFs (Instrumental Response Functions) version P6\_V3.

\section{Analysis method}

We monitor constantly at a daily basis the list of flares detected by 
RHESSI\cite{rhessi} and the \emph{Solar Monitor} web site\cite{solarmonitor}. We 
select the flares seen by RHESSI and GOES with more than $10^{5}$ counts 
(detected by RHESSI). For each of these flares we compute start and end time of 
the event in \emph{Fermi} MET (Mission Elapsed Time), the position of the Sun 
during the flare and the angle of the Sun direction with the LAT boresight.\\ For 
flares within the LAT field of view (angle with the LAT boresight $<80^{\circ}$) 
we search for excess of events in the LAT data. Although the Sun is a moving 
source in the sky, covering about $1^{\circ}$ per day, in this analysis we 
consider the Sun as a fixed source, due to the short duration of the flare events 
($<1$~h). As analysis method we use a likelihood fitting technique performed with 
a model that  includes the Sun as a point source and fixed galactic and 
extragalactic diffuse emission.\\ 

\section{Results}

\begin{figure}[!t]
  \centering
  \includegraphics[width=2.5in]{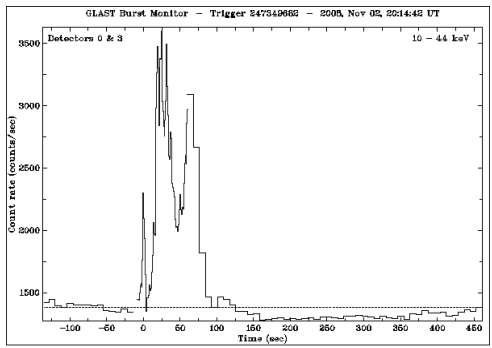}
  \caption{GBM light curve of the 2008 November 02 solar flare. Courtesy 
  of V. Connaughton (UAH).}
  \label{flare_GBM}
 \end{figure}

At 20:14:42.77 UT on 02 November 2008, \emph{Fermi}-GBM triggered and located a 
very soft and bright event\cite{gcn}. The event location was RA = 217.6 deg, Dec 
= -15.7 deg ($\pm 1.1$ deg), in excellent agreement with the Sun location. The 
time of the event coincides with the solar activity reported in the GOES solar 
reports (event 9790: onset at 20:12 UT, max at 20:15 UT, end at 20:17, B5.7 
flare). This is the first GBM detection of a solar flare. The GBM light curve 
(fig. \ref{flare_GBM}) shows a multiple peak event lasting approximately 177 s 
(8-30 keV). The event fluence (8-30 keV) in this time interval is $(1.54\pm 
0.03)\cdot 10^{-4}\textnormal{ erg}\textnormal{ cm}^{-2}$.\\ We selected LAT data 
in the energy range 100~MeV - 300~GeV between Nov 02 12:00 to 21:00 UT, according 
to the solar activity detected by GOES and RHESSI. No high energy emission has 
been detected by the LAT. 

\noindent From August 2008 to all May 2009 RHESSI has detected 150 flares with 
$>10^{5}$ counts. The highest energy band in which most of these flare have been 
observed by RHESSI is 3-6 keV. Few flares ($<20$) have been observed in the 
energy band 6-12 or 12-25 keV. We discarded flares outside the LAT field of view 
and the ones that occurred while the LAT was transiting in the SAA. As a result 
we analysed LAT data of about 90 flares computing upper limits on the high energy 
($>100\textnormal{ MeV}$) emission. 

\section{Conclusions}
We have searched for solar flare events in the first 10 months of LAT data 
(August 2008 - May 2009). Up till now we have no evidence of high energy emission 
from solar flares detected by the LAT, while the quiet Sun emission has been 
detected\cite{quietSun}\cite{moriond}\cite{AIP}. However, the Sun is at the 
minimum of its activity cycle and no intense flare has occurred. The solar 
activity is expected to rise in the next months, reaching the maximum in 2012. We 
will continue to monitor the active regions of the Sun and to improve our 
analysis techniques, waiting for an intense flare detectable by the LAT. 

\section{Acknowledgements} 
The \emph{Fermi} LAT Collaboration acknowledges support from a number of agencies 
and institutes for both development and the operation of the LAT as well as 
scientific data analysis. These include NASA and DOE in the United States, 
CEA/Irfu and IN2P3/CNRS in France, ASI and INFN in Italy, MEXT, KEK, and JAXA in 
Japan, and the K. A. Wallenberg Foundation, the Swedish Research Council and the 
National Space Board in Sweden. Additional support from INAF in Italy for science 
analysis during the operations phase is also gratefully acknowledged.


\begin{thebibliography}{99}
   \bibitem{Fermi} A.~Abdo \emph{et al.}, 2009, submitted to Astroparticle Physics, arXiv:0904.2226v1.
   \bibitem{LAT} W.~B.~Atwood \emph{et al.}, 2009, \emph{The Astrophysical Journal}, 697 1071.
   \bibitem{share} G.~Share, R.~Murphy, 2007, \emph{GLAST FIRST SYMPOSIUM}, AIP Conference Proceedings, 921.
   \bibitem{share06} G.~Share \emph{et al.}, 2006, \emph{BAAS}, 38, 255.
   \bibitem{grechnev} V.V.~Grechnev, 2008, \emph{Sol. Phys.}, 252, 149. 
   \bibitem{kanbach} G.~Kanbach \emph{et al.}, 1993, \emph{A\&AS}, 97, 349.
   \bibitem{nugget} L.~Svalgaard, 2009, \emph{RHESSI Science Nugget 99}.   
   \bibitem{rhessi} \emph{Rhessi flare list}, \url{http://hesperia.gsfc.nasa.gov/hessidata/dbase/hessi_flare_list.txt}.
   \bibitem{solarmonitor} \emph{Solar Monitor}, \url{http://www.solarmonitor.org/}. 
   \bibitem{gcn} C.~Kouveliotou, \emph{GCN Circular 8477}.
   \bibitem{quietSun} E.~Orlando \emph{Fermi-LAT Observation of quiescent solar emission}, these proceedings.
   \bibitem{moriond} M.~Brigida, 2009, 44th Rencontres de Moriond Proceedings.
   \bibitem{AIP} N.~Giglietto, 2009, AIP Conference Proceedings, 1112 238.


   \end{thebibliography}
\end{document}